\documentclass[doublecol]{epl2} 
\usepackage{subfigure}
\usepackage{color}
\def\be{\begin{equation}}
\def\ee{\end{equation}}

\def\bc{\begin{center}} 
\def\ec{\end{center}}
\def\bea{\begin{eqnarray}}
\def\eea{\end{eqnarray}}
\newcommand{\avg}[1]{\langle{#1}\rangle}

\title{Phase diagram  of the Bose-Hubbard Model  on  Complex Networks}
\shorttitle{Phase diagram  of the Bose-Hubbard Model  on  Complex Networks} %Insert here a short version of the title if it exceeds 70 characters

\author{Arda Halu\inst{1}, Luca Ferretti\inst{2}, Alessandro Vezzani \inst{3,4}, Ginestra Bianconi\inst{1}}
\shortauthor{A. Halu \etal}

\institute{                    
  \inst{1}   Department of Physics, Northeastern University, Boston, 
Massachusetts 02115 USA.\\
\inst{2} Centre de Recerca en AgriGenomica, Universitat Autonoma de Barcelona, 08193 Bellaterra, Spain\\\
\inst{3} Dipartimento di Fisica, Universit\'a degli Studi di Parma, V.le G.P. Usberti n.7/A, 43100 Parma, Italy\\
\inst{4} Centro S3, CNR Istituto di Nanoscienze, via Campi 213/a, 41100 Modena, Italy
}


\abstract{Critical phenomena can show unusual phase diagrams when defined in complex network topologies. The case of classical phase transitions such as the classical Ising model and the percolation transition has been studied extensively in the last decade. Here we show that the phase diagram of the Bose-Hubbard model, an exclusively quantum mechanical phase transition, also changes significantly when defined on random scale-free networks. We present a mean-field calculation of the model in annealed networks and we show that  when the second moment of the average degree diverges the Mott-insulator  { phase disappears in the thermodynamic limit. Moreover we study the model on quenched networks and we show that the Mott-insulator phase disappears in the thermodynamic limit as long as the maximal eigenvalue of the adjacency matrix diverges.} Finally we study the phase diagram of the model on Apollonian scale-free networks that can be embedded in 2 dimensions showing the extension of the results also to this case.
}

\begin{document}

\maketitle

\section{Introduction}
Recently great attention \cite{Dorogovtsev,Alain} has been addressed to critical phenomena unfolding on  complex networks. In this context it has been observed that the topology of the networks might significantly change the phase diagram of dynamical processes. 
For example when networks have a scale-free degree distribution $P(k)\sim k^{-\lambda}$ and the second moment $\avg{k^2}$ diverges with the network size, i.e. $\lambda\in(2,3]$ the Ising model \cite{ising1,ising2,ising3,Doro_expo}, the percolation phase transition \cite{percolation1, percolation2} and the epidemic spreading dynamics  {  on annealed networks}\cite{epidemics1} are strongly affected. Moreover  the spectral properties of the networks drive  { the epidemic spreading on quenched networks \cite{Durrett,epidemics2}}, the synchronization stability \cite{Synchr1,Synchr2}, the critical behavior of $O(N)$ models \cite{Burioni_ON1,Burioni_ON2} and the critical fluctuations of an Ising model on spatial scale-free networks \cite{Bradde}. Quantum critical phenomena also might depend on the topology of the underlying lattice as it has been shown for Bose-Einstein condensation in heterogeneous networks \cite{Burioni_BEC}.
Although large attention has been devoted to classical critical phenomena on scale-free networks, the behavior of quantum critical phenomena on scale-free networks has just started to be investigated.
{
In particular the  Anderson localization \cite{Havlin_localization_1,Havlin_localization_2} was studied in complex networks showing that by modulating the clustering coefficient of the network one might induce localization transition in scale-free networks.
Moreover, attention has been addressed to quantum processes on Apollonian networks \cite{
JSAndrade2005,RFSAndrade2005}, which provide an example of scale-free networks embedded in two dimensions. The quantum processes investigated are the Hubbard model \cite{Hubbard_Apollonian}, the free electron gas within  the tight-binding model \cite{Free_Electron_Gas_Apollonian}, and the topology induced Bose-Einstein condensation in Apollonian networks \cite{Bose_Einstein_Apollonian}. The study of quantum phase transitions on these networks has attracted attention particularly in recent years motivated by the creation of a new self-similar macromolecule -- a  nanometer-scale Sierpinski hexagonal gasket \cite{gasket}.
Recently \cite{QTIM} it has been shown that the Random Transverse Ising model is strongly affected by a scale-free network topology of the underlying networks on which it is defined and in particular by the second moment of the degree distribution $\avg{k^2}$. Indeed the critical temperature for the onset of the disordered phase is infinite if this second moment diverges and the network is scale-free with power-law exponent $\lambda\leq 3$.}

In this paper we investigate a critical process with no classical equivalent, the Bose-Hubbard model on complex networks \cite{Sachdev,Fisher, Vezzani, Vezzani2}. 
In the framework of ultracold atom physics nowadays defect free potentials have been constructed and the phenomenology of systems described by the Bose-Hubbard model experimentally reproduced \cite{Greiner}. By superimposing different lattices with incommensurate lattice constants, disordered systems of ultracold atoms have been recently experimentally investigated \cite{Inguscio}.
As the experimental technology advances it might become possible to investigate the role of topological network complexity in the phase diagram of the Bose-Hubbard model.
Moreover this complex topology might be effectively present in complex  granular materials and might affect insulator-superconductor phase transitions \cite{G} in these systems therefore a full account of the consequence of complex topologies might turn out to shed some light in the phase diagram of these complex materials.  {In fact the model defined on complex networks might provide a useful mean-field approximation to real disordered granular materials that captures essential features of their heterogeneity.
Finally the Hubbard model is a theoretical model that has applications far beyond condensed matter physics \cite{Hamma1,Hamma2} and investigating its properties on scale-free networks might 
stimulate further applications to other fields.}
Here we investigate this model when it is defined on a scale-free network topology.
Recently different approaches have been suggested for the theoretical study of the Bose-Hubbard model. Here we cite field theoretic approximations \cite{Fisher,Young,Svitsunov,Zimanyi}, mean-field approximations, \cite{Vezzani,Vezzani2,Shishedri,Bouchaud} quantum Monte Carlo simulations \cite{QMC1_1,QMC1_2,QMC1_3, QMC1_4,QMC1_5,QMC2}, quantum cavity methods \cite{Semerjian} among many others.

In this paper we characterize the phase diagram of the Bose-Hubbard model by mean-field approximation on annealed and quenched networks.
The Bose-Hubbard model is described by the Hamiltonian 
  \be 
\hat H=\sum_i \frac{U}{2} n_i(n_i-1)-\mu n_i-t \sum_{i,j}\tau_{ij}a_ia^{\dag}_j
\label{H0}\ee
where the indices $i,j=1,\ldots N$ indicate the nodes of the network. The network has an adjacency matrix ${ \bf \tau}$ such that $\tau_{ij}=1$ if there is a link between node $i$ and node $j$, otherwise $\tau_{ij}=0$. The operator $a_i^{\dag} (a_i)$ creates (annihilates) a boson at site $i$ and $n_i=a_i^{\dag}a_i$ counts the bosons at site $i$.
The parameter $U$ represents the repulsive boson-boson interaction, $t$ represents  the hopping amplitude between neighboring nodes {while $\mu$ indicates  the chemical potential}.
The hallmark of this Hamiltonian is a quantum phase transition between the Mott insulating phase and the superfluid phase originating from the competition  between the kinetic and the repulsive terms of the Hamiltonian.
We have to mention that the Mott insulating phase with vanishing compressibility is present strictly speaking only at zero temperature. At finite temperature thermal fluctuations induce a phase transition between the superfluid and normal phase.

In this paper we  show by  mean-field approximations that the phase diagram of the model   { defined on an annealed network}  depends on the second moment of the degree distribution $\avg{k^2}$. In particular, by the mean-field approximation, we found both for annealed and quenched networks that for scale-free networks with $\lambda\leq 3$ the Mott insulator phase reduces with increasing network size, disappearing in the thermodynamic limit.  {Moreover we observe differences between the model defined on a quenched network and an annealed network. In fact it is sufficient for a quenched random network to have diverging maximum degree in order to reduce the Mott-insulator phase to zero in the thermodynamic limit.} 
This demonstrates that complex  networks might strongly perturb the phase diagram of quantum phase transitions.

The paper is structured as follows: In section II we give the solution of the Bose-Hubbard model in annealed complex networks within the mean-field approximation. In section III we study numerically the phase diagram of the Bose-Hubbard model on quenched scale-free networks in the mean-field approximation and also on Apollonian networks. Finally we give the concluding remarks.

\section{Mean-field solution of the Bose-Hubbard Model on annealed complex networks}

An annealed network  evolves dynamically on the same time scale as the dynamical process occurring on it. During this process, links are created and annihilated but the expected degree of each node remains the same.
Solving dynamical models in annealed networks is usually straightforward. We are nevertheless in general not guaranteed that the phase diagram of the dynamical model on annealed scale-free networks will capture the essence of the dynamical model on quenched networks.
Our strategy here will be first to study the Bose-Hubbard model  in the annealed approximation and then to study the model on quenched networks to validate the main conclusions of the paper.  

We consider the ensemble of uncorrelated networks in which we assign to each node a hidden variable $\theta_i$ from a distribution $p(\theta)$ indicating the expected number of neighbors of a node. 
We consider the ensemble of networks for which the probability to draw a link between node $i$ and $j$ is given by  $p_{ij}$  
\be
p_{ij}=P(\tau_{ij}=1)=\frac{\theta_i \theta_j }{\avg{\theta} N}.
 \label{pij2}
\ee

In this ensemble the degree $k_i$ of a node $i$ is a Poisson random variable with expected degree $\overline{k_i}=\theta_i$. 
Therefore we will have 
\bea
\avg{\theta}&=&\overline{\avg{k}}\nonumber \\
\avg{\theta^2}&=&\overline{\avg{k(k-1)}}.
\label{uno}
\eea
where $\avg{\ldots}$ indicates the average over the $N$ nodes of the network and the overline in Eq. $(\ref{uno})$ indicates the average over  the ensemble of the networks.
We assume that the expected degree distribution of the network  ensemble is given by 
\bea
p(\theta)={\cal N} \theta^{-\lambda}e^{-\theta/\xi}
\label{exp}
\eea
where ${\cal N}$ is a normalization constant and $\xi $  is an exponential cut-off in the expected degree distribution.

In order to study the Bose-Hubbard model on annealed complex networks 
we consider  the fully connected Hamiltonian given by 
\bea 
 {H}=\sum_i \frac{U}{2} n_i(n_i-1)-\mu n_i-t \sum_{i,j}p_{ij}a_ia^{\dag}_j.
\eea
where in order to account for the dynamical nature of the annealed graph we have substituted the adjacency matrix element $\tau_{ij}$ in $H$ given by Eq. $(\ref{H0})$ with the matrix element $p_{ij}$ given by Eq. $\ref{pij2}$.
Moreover we perform the mean-field approximation to the Bose-Hubbard model introduced in  \cite{Shishedri} by taking
{
\bea
a_ia^{\dag}_j&\simeq&\avg{a_i}a^{\dag}_j+a_i\avg{a^{\dag}_j}-\avg{a_i}\avg{a^{\dag}_j}\nonumber \\
&\simeq & \psi_ia^{\dag}_j+a_i\psi_j-\psi_i\psi_j
\label{mf}
\eea}
where $\psi_i=\avg{a_i}=\avg{a^{\dag}_i}$ .
The Hamiltonian is then decomposed in single site terms
\begin{equation}
H=\sum_i H_i+{\avg{\theta}N}t\gamma^2
\end{equation}
with $H_i$ given by 
\begin{equation}
H_i=\frac{U}{2}n_i (n_i-1)-\mu n_i-t\theta_i \gamma (a_i+a^{\dag}_i)
\label{H}
\end{equation}
and with  $\gamma$  indicating the order parameter of the superfluid phase, defined as
\begin{equation}
\gamma=\frac{1}{\avg{\theta}N}\sum_i \theta_i \psi_i.
\label{gamma}
\end{equation}
In this mean-field picture the Hamiltonian therefore decouples in single site (node) Hamiltonians $H_i$ depending on the mean field order parameter $\gamma$.
We can therefore write the single site (node) Hamiltonian as an unperturbed Hamiltonian plus an interaction depending on the parameter $\gamma$, i.e.
\begin{equation}
H_i=H^{(0)}_i+\gamma  \theta_i V_i
\end{equation}
with 
\begin{eqnarray}
H^{(0)}_i&=&\frac{U}{2}n_i (n_i-1)-\mu n_i\nonumber \\
V_i&=&t (a_i+a^{\dag}_i)
\end{eqnarray}
The ground state energy $E^{(0)}_i(n)=E^{(0)}(n)$ with
$E^{(0)}(n^{\star})=0$ if   $\mu<0$ and $E^{(0)}(n^{\star})=
-\mu n^{\star}+\frac{1}{2}U n^{\star}(n^{\star}-1)$ if  $\mu \in (U(n^{\star}-1),Un^{\star})$
The second order correction to the energy is given by $E^{(2)}_i$
\begin{eqnarray}
E^{(2)}_i(n^{\star})&=&\gamma^2 \theta_i^2 \sum_{n \neq n^{\star} }\frac{|\avg{n|V_i|n^{\star}}|^2}{E^{(0)}(n^{\star})-E^{(0)}(n)}\nonumber \\
&=& \gamma^2 t^2 \theta_i^2 \left(\frac{n^{\star}}{U(n^{\star}-1)-\mu}+\frac{n^{\star}+1}{\mu-Un^{\star}}\right)\nonumber
\end{eqnarray}
Therefore the energy spectrum $E$ is given by the eigenvalues of the Hamiltonian $H$ Eq. (\ref{H}), i.e.
\begin{equation}
E=\mbox{const} +m^2 \gamma^2
\end{equation}
with 
\begin{equation}
\frac{m^2}{t\avg{\theta} N}=1+t \frac{\avg{\theta^2}}{\avg{\theta}}\left(\frac{n^{\star}}{U(n^{\star}-1)-\mu}+\frac{n^{\star}+1}{\mu-Un^{\star}}\right).
\end{equation}
The phase transition between a Mott-insulator phase where $\gamma=0$ and a superfluid phase where $\gamma>0$ occurs when $m=0$. Therefore the phase diagram at $T=0$ is given by 
\begin{equation}
t_c(U,\mu,T=0)=U\frac{\avg{\theta}}{\avg{\theta^2}}\frac{[\mu/U-n^{\star}][(n^{\star}-1)-\mu/U]}{\mu/U+1}
\end{equation}
with $\mu/U\in[n^{\star}-1,n^{\star}]$.
The difference with respect to the mean-field phase diagram for regular lattices is that $t_c$, the critical hopping rate, depends on the second moment of the expected degree distribution, i.e.  $\avg{\theta^2}=\avg{k(k-1)}$. Given the  general expression for the expected degree distribution of complex networks considered in this paper Eq. $(\ref{exp})$ including the exponential cut-off $\xi$,  the Mott-insulator phase disappears as $\xi \to\infty $ when $\lambda\leq3$ and remains finite instead if $\lambda>3$ and $\xi\to\infty$.
Therefore as $\frac{\avg{\theta^2}}{\avg{\theta}}$ diverges i.e. as $\xi \to \infty $ while $\lambda\leq 3$ we have that the Mott insulator phase shrinks and finally disappears for large network sizes. Also it can be seen that the critical indices will deviate from the mean-field values and they can be found by applying the heterogeneous mean-field techniques \cite{Doro_expo}
 developed for the classical phase transition.

At finite temperature we cannot properly speak about a Mott insulator phase but we have still a phase diagram between a normal phase and the superfluid phase.
The local order parameter is given by the thermal average of the creation and annihilation operators, i.e.
\begin{equation}
\psi_i=\frac{\mbox{Tr} a_i e^{-\beta H_i}}{\mbox{Tr} e^{-\beta H_i}}.
\end{equation}
%and it  can be easily shown that superfuid order parameter $\gamma$ given by Eq. $(\ref{gamma})$ can be written as 
%\begin{equation}
% \gamma= \frac{1}{2\avg{\theta}N} \sum_i \frac{\partial \log Z_i(\beta)}{\partial \gamma}
%\end{equation} 
%where $Z_i(\beta)= \mbox{Tr} e^{-\beta H_i}$.
Using the same steps as in \cite{Vezzani} we can prove that the critical line for the Mott-insulator, superfluid phase is given by
\begin{equation}
t_c(U,\mu,\beta)=\frac{\avg{\theta}}{\avg{\theta^2}}\frac{\sum_{r=0}^{\infty}e^{\beta[\mu r-(U/2)r(r-1)]}}{\sum_{r=0}^{\infty}Q_r(U,\mu)e^{\beta[\mu r-(U/2)r(r-1)]}}
\end{equation}
where
\begin{equation}
Q_r(U,\mu)=\frac{\mu+U}{(\mu-Ur)(U(r-1)-\mu)}.
\end{equation}
Therefore the phase diagram at finite temperature is also affected by the topology of the network and significantly changes when $\avg{\theta^2}$ diverges.
\begin{figure}
\begin{center}
\includegraphics[width=.75\columnwidth]{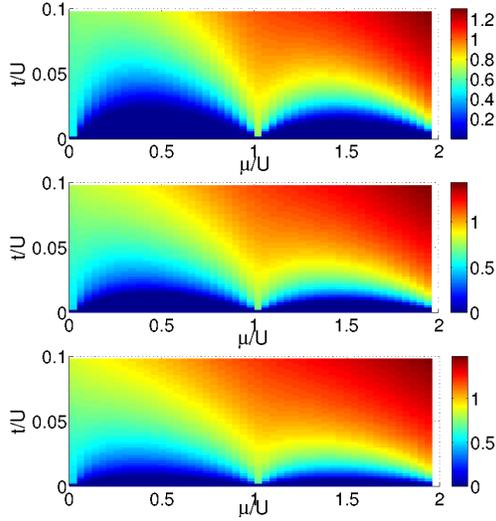}
\end{center}
\caption{(Color online)   Average order parameter for the superfluid phase for scale-free networks with power-law exponent $\lambda=2.2$ and network sizes $N=100$ (top)$N=1000$ (middle) and $N=10,000$ (bottom). As the network size increases the phase diagram changes monotonically as predicted by the mean-field treatment in the case $\lambda<3$. Therefore there is no Mott-insulator phase in the limit $N\to \infty$.  }
\label{figure_2.2}
\end{figure}\begin{figure}
\begin{center}
\includegraphics[width=.75\columnwidth]{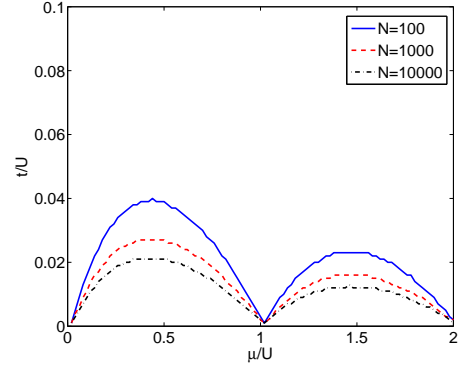}
\end{center}
\caption{(Color online)     Average order parameter for the superfluid phase for scale-free networks with power-law exponent $\lambda=3.5$ and network sizes $N=100$, $N=1000$ and $N=10,000$. As the network size increases the phase diagram has slower finite size effects with respect to the case $\lambda<3$. Nevertheless the maximal eigenvalue increases with network size as $\Lambda\propto \sqrt{k_{max}}$ and therefore the Mott Insulator phase disappears in the thermodynamic limit also in this case. }
\label{figure_3.5}
\end{figure}

\begin{figure}
\centering
\mbox{\subfigure{\includegraphics[width=0.45\columnwidth]{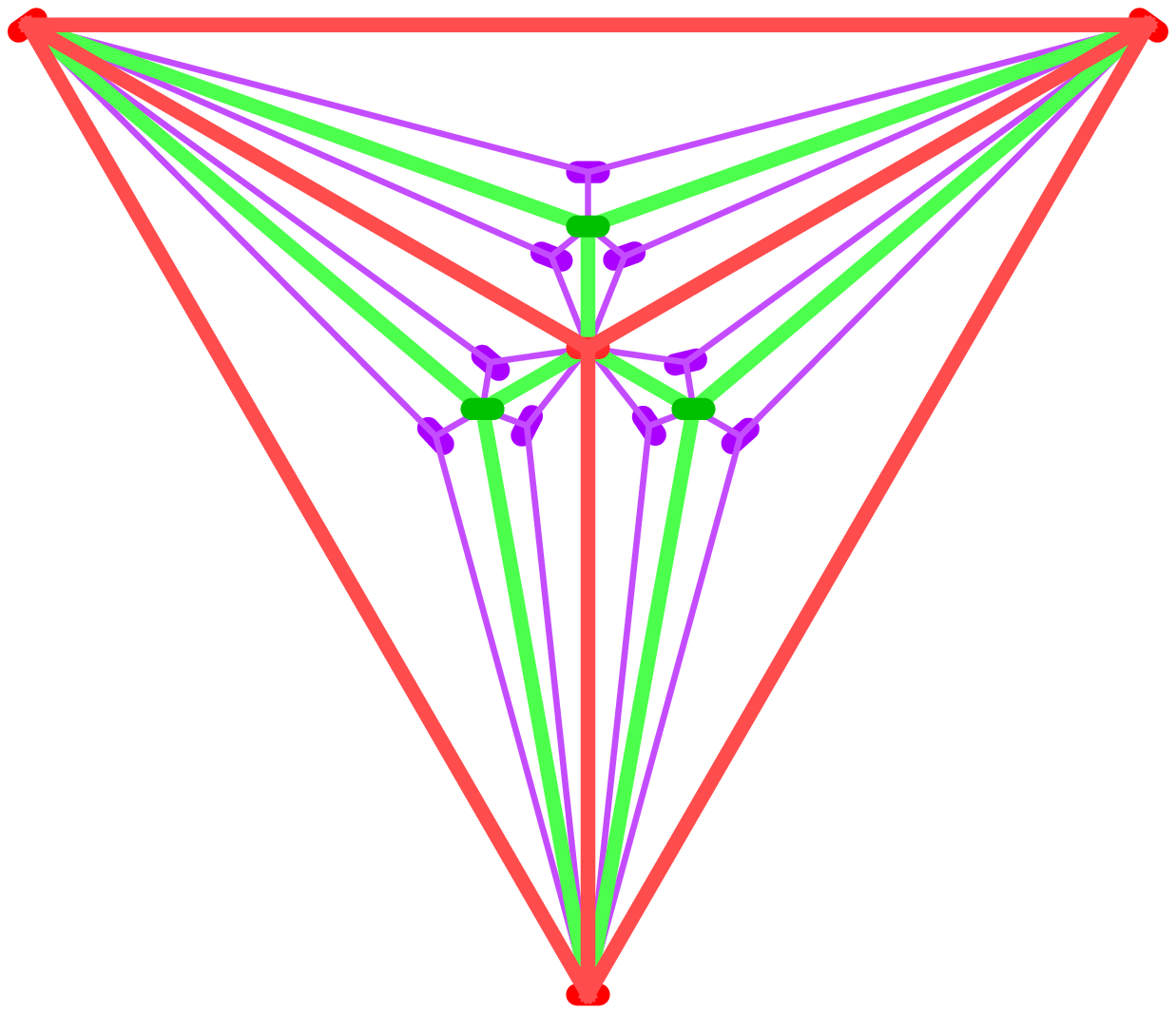}}\quad
\subfigure{\includegraphics[width=0.45\columnwidth]{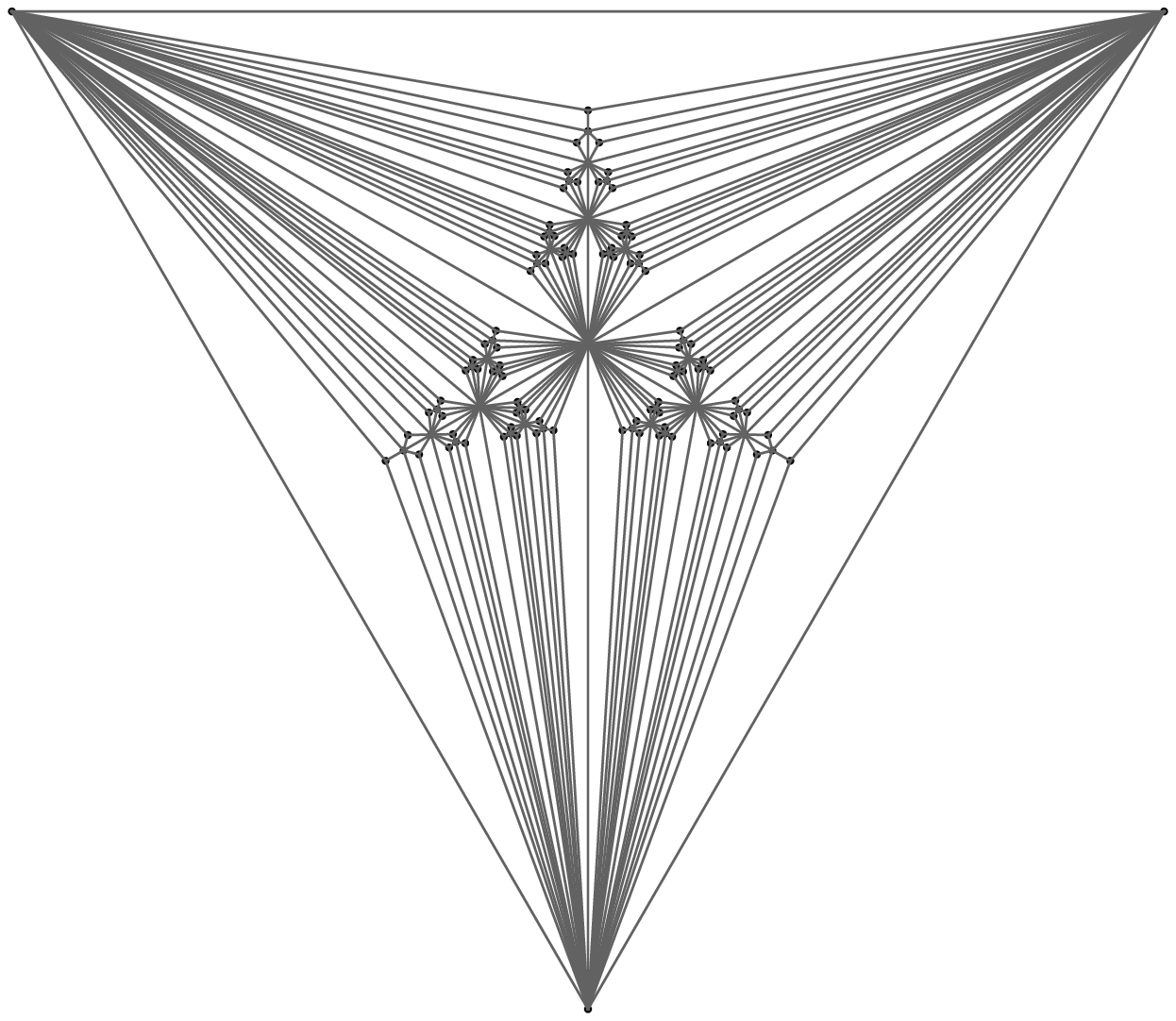}}}
\caption{(Color online)  Left panel: the first 3 generations of the Apollonian graph. The nodes and links added to construct the 1st (red), 2nd (green) and 3rd (purple) generations are shown. Right panel: the 5th generation Apollonian network.}
\label{apollonian}
\end{figure}

\begin{figure}
\begin{center}
\includegraphics[width=0.9\columnwidth]{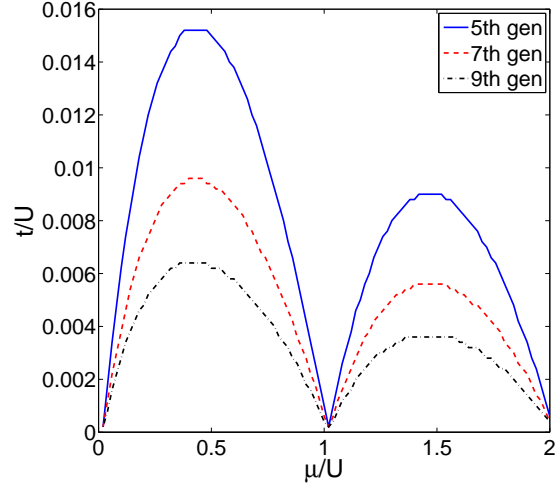}
\end{center}
\caption{(Color online) Effective phase diagram of the Bose-Hubbard model on Apollonian networks of  the 5th, 7th and 9th generation, with network sizes $N=124$, $N=1096$ and $N=9844$, respectively. As the network size diverges the Mott-Insulator phase is reduced and it disappears in the limit of an infinite network.}
\label{apollophase}
\end{figure}
\section{Phase diagram of the Bose-Hubbard phase transition on quenched complex networks}

We now discuss the phase diagram of the Bose-Hubbard model on quenched networks.
In particular we focus on the base diagram at $T=0$.
The Hamiltonian we consider is the original Bose-Hubbard Hamiltonian defined on the adjacency matrix ${\bf \tau}$ {(Eq. $(\ref{H0})$).} 
%  \be 
%\hat H=\sum_i \frac{U}{2} n_i(n_i-1)-\mu n_i-t \sum_{i,j}\tau_{ij}a_ia^{\dag}_j
%\label{H1}\ee
We solve this equation again in the mean-field approximation assuming {Eq. $(\ref{mf})$}
%\begin{equation}
%a_ia^{\dag}_j\simeq\avg{a_i}a^{\dag}_j+a_i\avg{a^{\dag}_j}-\avg{a_i}\avg{a^{\dag}_j}.
%\end{equation}
and  $\psi_i=\avg{a_i}=\avg{a^{\dag}_i}$.
By solving self-consistently for $\psi_i$ we found the boundary of the Mott-insulator phase where $\psi_i=0$ and the superfluid phase where $\psi_i>0 \forall i$.
{The mean-field Hamiltonian $H^{MF}$ is then parametrized by the self-consistent parameters $\psi_i$ and reads
\bea
H^{MF}&=&\sum_i \left[\frac{U}{2}n_i (n_i-1)-\mu n_i-t\sum_j \tau_{ij} (a_i+a^{\dag}_i)\psi_j\right]\nonumber \\
&&+t \sum_i\sum_j \tau_{ij} \psi_i \psi_j.
\eea}
Following \cite{Vezzani2} we consider the hopping term as a perturbation. At the first order of the perturbation theory  we get that
\begin{equation}
\psi_i=\frac{t}{U} F(\mu,U)\sum_{j} \tau_{ij} \psi_j
\end{equation}
where 
\begin{equation}
F(\mu,U)=\frac{\mu+U}{[\mu-n^{\star}U][U(n^{\star}-1)-\mu]}
\end{equation}
where $\mu\in{(U(n^{\star}-1),U n^{\star})}$.
Therefore, if $\Lambda$ is the maximal eigenvalue of the adjacency matrix $\{\tau_{ij}\}$, the  Mott-insulator phase with $\psi_i=0$ is stable as long as  
\begin{equation}
\frac{t}{U}F(\mu,U) \Lambda<1.
\end{equation} 
We observe that in random scale-free networks with degree distribution $p(k)={\cal N}k^{-\lambda}$  {  the maximal eigenvalue $\Lambda$ of the adjacency matrix diverges with a diverging value of the maximal degree of the network $k_{max}$ as $\Lambda\propto \sqrt{k_{max}}$} \cite{doro_spectra, Others, Chung, K}.
 {Therefore we find that also as long as the maximal degree of the network diverges the Mott-insulator phase disappears in the large network limit, changing the phase diagram of the model significantly with respect to regular networks where the maximal degree of the network remains constant.}
We have checked these results by performing numerical integration of the mean-field calculations. We have studied  the phase diagram of single quenched networks with scale-free degree distribution $p(k)={\cal N}k^{-\lambda}$ and  { different values of the power-law exponent $\lambda$ to see how fast the convergence of the solution to the asymptotic phase diagram is.}
In the following we will show our  finite-size scaling calculations and the resulting  effective phase diagram of the Bose-Hubbard model within the mean-field approximation on the quenched network for different values of the number of nodes $N$.
In figure $\ref{figure_2.2}$ we plot the effective phase diagram for network sizes $N=100,1000,10000$ finding that for $\lambda=2.2<3$ the boundary of the Mott-insulating phase decreases with the network  size. On the other hand for a typical network with $\lambda=3.5>3$ (see figure $\ref{figure_3.5}$) the phase diagram has  { slower} finite size dependencies.
This shows that the annealed  approximation for the Bose-Hubbard model on scale-free networks  {strongly differs from the quenched phase diagram of the model for $\gamma>3$.
In particular we have that the Mott insulator phase transition on annealed scale-free networks, with diverging second moment of the degree distribution, vanishes in the thermodynamic limit while in quenched networks it is sufficient that the most connected node has a diverging connectivity in the thermodynamic limit to destroy the Mott insulator phase}.

Finally we have studied the phase diagram as predicted by the mean-field approximation, on Apollonian networks \cite{JSAndrade2005}.
Apollonian networks are an example of scale-free networks which are embedded in a two dimensional space. They are constructed by  considering  the classical 2D Apollonian packing model in which the space between three tangent circles placed on the vertices of an equilateral triangle is filled by a maximal circle. The space-filling procedure is repeated for every space bounded by three of the previously drawn tangent circles. The corresponding Apollonian network is constructed by connecting the centers of all the touching circles (Fig. \ref{apollonian} (left)).
The resulting network is scale-free with power-law degree distribution $p(k)={\cal N}k^{-\lambda}$ and $\lambda=1+\ln(3)/ln(2)\simeq 2.585$. Also these networks are known to have diverging maximal eigenvalue $\Lambda$ of their adjacency matrix \cite{RFSAndrade2005}. Therefore we expect that also in these networks the Mott-insulator phase of the Bose-Hubbard model should disappear in the large network limit.
 We have used the iterative scheme introduced in \cite{RFSAndrade2005} to generate 5th (Fig. \ref{apollonian} (right)), 7th and 9th generation Apollonian networks, where the number of nodes in the $n$th generation is $N(n) = (3^n + 5)/2$. In Fig. \ref{apollophase} we demonstrate the finite-size effects on the phase diagram found in the mean-field approximation.
 {We note here that in experimental  realizations of the Bose-Hubbard model on Apollonian networks it might be relevant to include a hopping term that depends also on the distance between the nodes. The dynamics on this weighted network might further modify the phase diagram of the process. In this paper we have chosen to study the simple case in which the hopping doesn't depend on the distance, leaving the  characterization of the dynamics on the  weighted network for future investigations.}

\section{Conclusions}
In conclusion in this paper we have shown that the scale-free network topology of the underlying network strongly affects the phase diagram of the Bose-Hubbard model.
By performing mean-field calculations on annealed networks we have shown that the Mott-insulator phase disappears in the large network limit as long as the power-law exponent $\lambda$ of the degree distribution is $\lambda\leq3$ and the exponential cutoff $\xi$ of the distribution diverges. We have performed mean-field calculations in annealed networks finding in this approximation the phase diagram of the model both at $T=0$ and at finite temperature. Moreover  the  analytical and numerical solutions of the Bose-Hubbard model on quenched networks show  {that this argument must be corrected in the quenched case and that it is sufficient that the maximal eigenvalue diverges in order to change the phase diagram of the model}. Finally we have considered the Bose-Hubbard model on Apollonian networks that are an example of scale-free networks embedded in a two dimensional space.
In short, { this work offers a new perspective on the characterization of  quantum critical phenomena in annealed and quenched complex networks and shows that the second moment of the degree distribution $\avg{k^2}$ and  { the maximal eigenvalue of the adjacency matrix  play} a crucial role in determining the phase diagram of the Bose-Hubbard model}.

\end{document}